\documentstyle[epsfig,feynmp]{aipproc}  

\newcommand{\absatz}{\vspace{2ex}\noindent}

\newcommand{\journal}[4]{{#1} \textbf{{#2}}, #3 (#4)}

\newcommand{\NPA}{\emph{Nucl.\ Phys.\ }{A}}
\newcommand{\NPB}{\emph{Nucl.\ Phys.\ }{B}}
\newcommand{\PLB}{\emph{Phys.\  Lett.\ }{B}}

\newcommand{\PRL}{\emph{Phys.\ Rev.\ Lett.\ }}

\newcommand{\dis}{\displaystyle}
\newcommand{\non}{\nonumber}

\newcommand{\ii}{{\rm i}}

\newcommand{\tr}{{\rm tr}}
\newcommand{\T}{{\rm T}}

\newcommand{\pv}{\vec{\,\!p}\!\:{}}

\newcommand{\mpi}{m_\pi}
\newcommand{\fpi}{f_\pi}
\newcommand{\MeV}{{\rm MeV}}
\renewcommand{\fm}{{\rm fm}}

\newcommand{\de}{\partial}
\newcommand{\dev}{\vec{\de}}

\begin{document}

\begin{fmffile}{coifeyn}
  \fmfset{curly_len}{2mm} \fmfset{dash_len}{1.5mm} \fmfset{wiggly_len}{3mm}
  \newcommand{\feynbox}[2]{\mbox{\parbox{#1}{#2}}}
  \renewcommand{\fs}{\scriptstyle}
  \newcommand{\hq}{\hspace{0.5em}} \newcommand{\hqm}{\hspace{-0.25em}}
  
  \fmfcmd{vardef ellipseraw (expr p, ang) = save radx; numeric radx; radx=6/10
    length p; save rady; numeric rady; rady=3/10 length p; pair center;
    center:=point 1/2 length(p) of p; save t; transform t; t:=identity xscaled
    (2*radx*h) yscaled (2*rady*h) rotated (ang + angle direction length(p)/2
    of p) shifted center; fullcircle transformed t enddef;
    style_def ellipse expr p= shadedraw ellipseraw (p,0); enddef; }


  \title{A Sketch of Two and Three Bodies\thanks{Talk held at the
      International Workshop on Hadron Physics ``Effective Theories of Low
      Energy QCD'' in Coimbra, Portugal, 10th -- 15th September 1999; to be
      published in the Proceedings; preprint numbers nucl-th/9911035,
      NT@UW-99-62, TUM-T39-99-25.}}
  
  \author{Harald W.~Grie\3hammer${}^{\dagger}$\thanks{Email:
      hgrie@physik.tu-muenchen.de}} \address{${}^{\dagger}$Nuclear Theory
    Group,
    Department of Physics, University of Washington,\\
    Box 351 560, Seattle, WA 98195-1560, USA\\ and \\
    Institut f{\"u}r Theoretische Physik, Physik-Department der\\
    Technischen Universit{\"a}t M{\"u}nchen, 85748 Garching, Germany
    (permanent address)}

  \maketitle

\begin{abstract}
  A cartoon of the Effective Field Theory of many nucleon systems is drawn,
  concentrating on Compton scattering in the two nucleon system, and on $nd$
  scattering in the three body system.
\end{abstract}

%

The purpose of this presentation is to give a concise introduction into the
Effective Field Theory (EFT, for a review see e.g.~\cite{LepageQEDlecture}) of
two and three nucleon systems as it emerged in the last three years.  However,
I can only give a ``teaser'' with a lot of words and figures and a few cheats
in details, referring to the literature, esp.~to the excellent proceedings of
the INT-Caltech Workshops 1998 and 1999~\cite{INTWorkshopSummary}. I
concentrate on work undertaken with J.-W.~Chen, R.P.~Springer and M.J.~Savage
in \cite{pola,Compton}, P.F.~Bedaque in \cite{pbhg,pbfghg}, and F.~Gabbiani in
\cite{pbfghg}. M.~Birse's talk at this Workshop provides a more formal
investigation of the EFT of the two nucleon system, and U.-G.~Mei\3ner's
alternative approach he presented here follows Weinberg's original
suggestion~\cite{Weinberg} but needs to be studied further.

\absatz Effective Field Theory methods are largely used in many branches of
physics where a separation of scales exists. In low energy nuclear systems,
the two well separated scales are, on one side, the low scales of the typical
momentum of the process considered and the pion mass, and on the other side
the higher scales associated with chiral symmetry and confinement. This
separation of scales was explored with great success in the mesonic sector
(Chiral Perturbation Theory~\cite{leutWeinberg79}) and in the one baryon
sector (Heavy Baryon Chiral Perturbation
Theory~\cite{GasseretalJenkinsManohar}), producing a low energy expansion of a
variety of observables (see also the Chiral Perturbation Theory section of
this Workshop). It provided for the first time a description of strongly
interacting particles which is systematic, rigorous and model independent
(meaning, independent of assumptions about the non-perturbative QCD dynamics).

\absatz Three main ingredients enter the construction of an EFT: The
Lagrangean, the power counting and a regularisation scheme.  First, the
relevant degrees of freedom have to be identified. In his original suggestion
how to extend EFT methods to systems containing two or more nucleons,
Weinberg~\cite{Weinberg} noticed that below the $\Delta$ production scale,
only nucleons and pions need to be retained as the infrared relevant degrees
of freedom of low energy QCD. Because at these scales the momenta of the
nucleons are small compared to their rest mass, the theory becomes
non-relativistic at leading order in the velocity expansion, with relativistic
corrections systematically included at higher orders. The most general
chirally (and iso-spin) invariant Lagrangean consists hence of contact
interactions between non-relativistic nucleons, and between nucleons and
pions, with the first few terms of the form
\begin{eqnarray}\label{ksw}
   {\mathcal{L}}_{NN}&=&N^{\dagger}(\ii \de_0+\frac{\dev^2}{2M})N+
   \;\frac{\fpi^2}{8}\;
   \tr[(\partial_{\mu} \Sigma^{\dagger})( \partial^{\mu} \Sigma)]\;+
   \;g_A N^{\dagger} \vec{A}\cdot\sigma N\;-\non\\
   &&-\;C_0 (N^{\T} P^i N)^{\dagger} \ (N^{\T} P^i N)\;
   +\\&&
   + \;\frac{C_2}{8}
   \left[(N^{\T} P^i N)^{\dagger} (N^{\T} P^i
     (\stackrel{\scriptscriptstyle\rightarrow}{\de}-
      \stackrel{\scriptscriptstyle\leftarrow}{\de})^2 N)+
   {\rm h.c.}\right]
   + \dots\;\;,\nonumber
\end{eqnarray}
where $N={p\choose n}$ is the nucleon doublet of two-component spinors and
$P^i$ is the projector onto the iso-scalar-vector channel, $
P^{i,\,b\beta}_{a\alpha}=\frac{1}{\sqrt{8}}
(\sigma_2\sigma^i)_{\alpha}^{\beta} (\tau_2)_a^b$.  $\sigma$ ($\tau$) are the
Pauli matrices acting in spin (iso-spin) space. The iso-vector-scalar part of
the $NN$ Lagrangean introduces more constants $C_i$ and interactions and has
not been displayed for convenience. The field $\xi$ describes the pion,
$\xi(x)=\sqrt{\Sigma}=e^{\ii \Pi/\fpi}$, $\fpi=130\;\MeV$. $D_{\mu}$ is
the chiral covariant derivative $D_{\mu}=\partial_{\mu}+V_{\mu}$, and the
vector and axial currents are $
V_{\mu}=\frac{1}{2}(\xi\partial_{\mu}\xi^{\dagger}+\xi^{\dagger}
\partial_{\mu}\xi)$, $
A_{\mu}=\frac{\ii}{2}(\xi\partial_{\mu}\xi^{\dagger}-\xi^{\dagger}
\partial_{\mu}\xi)$. The interactions involving pions are severely restricted
by chiral invariance. As such, the theory is an extension to the many nucleon
system of Chiral Perturbation Theory and Heavy Baryon Chiral Perturbation
Theory. Like in its cousins, all short distance physics -- branes and strings,
quarks and gluons, resonances like the $\Delta$ or $\sigma$ -- is integrated
out into the coefficients of the low energy Lagrangean. In principle, these
constants could be derived by solving QCD or via models of the short distance
physics like resonance saturation. The most common and practical way to
determine those constants, though, is by fitting them to experiment.

The EFT with pions integrated out (formally, $g_A=0$ in (\ref{ksw})) is valid
below the pion cut and was recently pushed to very high orders in the
two-nucleon sector~\cite{CRS} where accuracies of the order of $1\%$ were
obtained. It can be viewed as a systematisation of Effective Range Theory
with the inclusion of relativistic and short distance effects traditionally
left out in that approach.

\absatz Because the Lagrangean (\ref{ksw}) consists of infinitely many terms
only restricted by symmetry, an EFT may at first sight
suffer from lack of predictive power. Indeed, as the second part of its
formulation, predictive power is ensured only by establishing a power counting
scheme, i.e.\ a way to determine at which order in a momentum expansion
different contributions will appear, and keeping only and all the terms up to
a given order. The dimensionless, small parameter on which the expansion is
based is the typical momentum $Q$ of the process in units of the scale
$\Lambda$ at which the theory is expected to break down, with estimates
ranging from $\Lambda_{\pi}\approx 300$ to
$800\;\MeV$~\cite{INTWorkshopSummary} in the two body system for the theory
with pions. The pion-less theory should be in disagreement with experiment
starting at the pion cut, $\Lambda_{{\rm no\pi}}\approx 140\;\MeV$. Values for
$\Lambda$ and $Q$ have to be determined from
comparison to experiments and are a priori unknown. Assuming that all
contributions are of natural size, i.e.\ ordered by powers of $Q$, the
systematic power counting ensures that the sum of all terms left out when
calculating to a certain order in $Q$ is smaller than the last order retained,
allowing for an error estimate of the final result.

Even if calculations of nuclear properties were possible starting from the
underlying QCD Lagrangean, EFT simplifies the problem considerably by
factorising it into a short distance part (subsumed into the coefficient of
the Lagrangean) and a long distance part which contains the infrared-relevant
physics and is dealt with by EFT methods. EFT provides an answer of finite
accuracy because higher order corrections are systematically calculable and
suppressed in powers of $Q$. Hence, the power counting allows for an error
estimate of the final result, with the natural size of all neglected terms
known to be of higher order. Relativistic effects, chiral dynamics and
external currents are included systematically, and extensions to include
e.g.~parity violating effects are straightforward. Gauged interactions and
exchange currents are unambiguous.  Results obtained with EFT are easily
dissected for the relative importance of the various terms.  Because only
$S$-matrix elements between on-shell states are observables, ambiguities
nesting in ``off-shell effects'' are absent.  On the other hand, because only
symmetry considerations enter the construction of the Lagrangean, EFTs are
less restrictive as no assumption about the underlying QCD dynamics is
incorporated.

In systems involving two or more nucleons, establishing such a power counting
is complicated by the fact that unnaturally large scales have to be
accommodated, so that some coefficients in the Lagrangean may not be of
natural size and hence possibly jeopardise power counting: Given that the
typical low energy scale in the problem should be the mass of the pion as the
lightest particle emerging from QCD, fine tuning seems to be required to
produce the large scattering lengths in the ${}^1{\rm S}_0$ and
${}^3{\rm S}_1$ channels ($1/a^{{}^1{\rm S}_0}=-8.3\;\MeV,\;
1/a^{{}^3{\rm S}_1}=36\;\MeV$). Since there is a bound state in the
${}^3{\rm S}_1$ channel with a binding energy $B=2.225\;\MeV$ and hence a
typical binding momentum $\gamma=\sqrt{M B}\simeq 46\;\MeV$ well below the
scale $\Lambda$ at which the theory should break down, it is also clear that
at least some processes have to be treated non-perturbatively in order to
accommodate the deuteron. Most likely, these small scales do not arise from
the fact that the real world is close to the chiral limit: In the singlet
channel, for instance, the one pion exchange potential vanishes in the chiral
limit and thus cannot be the cause of the fine tuning. The fine tuning then
must be a result of short distance physics.

A way to incorporate this fine tuning into the power counting was suggested by
Kaplan, Savage and Wise \cite{KSW}: At very low momenta, contact interactions
with several derivatives -- like $p^2C_2$ and the pion-nucleon interactions --
should become unimportant, and we are left only with the contact interactions
proportional to $C_0$. The leading order contribution to nucleons scattering
in an ${\rm S}$ wave comes hence from four nucleon contact interactions and
is summed geometrically as in Fig.~\ref{fig:deuteronprop} to all orders to
produce the shallow real bound state, i.e.~the deuteron.
\begin{figure}[!htb]
  \begin{center}
    $
        \feynbox{25\unitlength}{
            \begin{fmfgraph*}(25,25)
              \fmfleft{i} \fmfright{o} \fmf{double,width=thin}{i,o}
            \end{fmfgraph*}}
          \hq=\hq\feynbox{30\unitlength}{
            \begin{fmfgraph*}(30,25)
              \fmfleft{i1,i2} \fmfright{o1,o2}
              \fmf{vanilla,width=thin}{i1,v,o2}
              \fmf{vanilla,width=thin}{i2,v,o1}
            \end{fmfgraph*}}  \hq+\hq
          \feynbox{45\unitlength}{
            \begin{fmfgraph*}(45,25)
              \fmfleft{i1,i2} \fmfright{o1,o2}
              \fmf{vanilla,width=thin,tension=6}{i1,v1}
              \fmf{vanilla,width=thin,tension=6}{v2,o2}
              \fmf{vanilla,width=thin,tension=6}{i2,v1}
              \fmf{vanilla,width=thin,tension=6}{v2,o1}
              \fmf{vanilla,width=thin,left=0.5}{v1,v2}
              \fmf{vanilla,width=thin,left=0.5}{v2,v1}
            \end{fmfgraph*}}  \hq+\hq
          \feynbox{63\unitlength}{
            \begin{fmfgraph*}(63,25)
              \fmfleft{i1,i2} \fmfright{o1,o2}
              \fmf{vanilla,width=thin,tension=6}{i1,v1}
              \fmf{vanilla,width=thin,tension=6}{v2,o2}
              \fmf{vanilla,width=thin,tension=6}{i2,v1}
              \fmf{vanilla,width=thin,tension=6}{v2,o1}
              \fmf{vanilla,width=thin,left=0.7}{v1,v3}
              \fmf{vanilla,width=thin,left=0.7}{v3,v1}
              \fmf{vanilla,width=thin,left=0.7}{v2,v3}
              \fmf{vanilla,width=thin,left=0.7}{v3,v2}
            \end{fmfgraph*}}  +\;\dots
     \;=\;\dis\frac{-C_0}{1-\rule{0pt}{12pt}\feynbox{24\unitlength}{
            \begin{fmfgraph*}(25,12)
              \fmfleft{i} \fmfright{o} \fmf{vanilla,width=thin,left=0.5}{i,o}
              \fmf{vanilla,width=thin,left=0.5}{o,i}
            \end{fmfgraph*}}}
        $
\vspace{10pt}
    \caption{Re-summation of the contact interactions into the deuteron
      propagator.}
    \label{fig:deuteronprop}
  \end{center}
\end{figure}

How to justify this? Any diagram can be estimated by scaling momenta by a
factor of $Q$ and non-relativistic kinetic energies by a factor of $Q^2/M$.
The remaining integral includes no dimensions and is taken to be of the order
$Q^0$ and of natural size. This scaling implies the rule that nucleon
propagators contribute one power of $M/Q^2$ and each loop a power of $Q^5/M$.
Assuming that
\begin{eqnarray}\label{scalingksw}
  C_0\sim\frac{1}{M Q}&\;\;,\;\;&
  C_2\sim\frac{1}{M \Lambda Q^2}\;\;,
\end{eqnarray}
the diagrams contributing at leading order to the deuteron propagator are
indeed an infinite number as shown in Fig.~\ref{fig:deuteronprop}, each one of
the order $1/(MQ)$. The regulator dependent, linear divergence in each of the
bubble diagrams does not show in dimensional regularisation as a pole in $4$
dimensions, but it does appear as a pole in $3$ dimensions which we subtract
following the Power Divergence Subtraction scheme~\cite{KSW}. Dimensional
regularisation is chosen to explicitly preserve the systematic power counting
as well as all symmetries (esp.~chiral invariance) at each order in every step
of the calculation. At leading (LO), next-to-leading order (NLO) and often
even NNLO in the two nucleon system, it also allows for simple, closed answers
whose analytic structure is readily asserted. The deuteron propagator
\begin{equation}
  \label{deuteronpropagator}
  \frac{4\pi}{M}\;\frac{-\ii}{\frac{4\pi}{MC_0}+\mu-
    \sqrt{\frac{\pv^2}{4}-Mp_0-\ii\varepsilon}}
\end{equation}
has the correct pole position and cut structure when one chooses
\begin{equation}
  \label{fixC0}
  C_0(\mu)=\frac{4\pi}{M}\;\frac{1}{\gamma-\mu}\;\;.
\end{equation}
Indeed, when choosing $\mu\sim Q$, the leading order contact interaction
scales as demanded in (\ref{scalingksw}) and -- as expected for a physical
observable -- the $NN$ scattering amplitude becomes independent of $\mu$, the
renormalisation scale or cut-off chosen. The same can be shown for the higher
order coefficients, so that the scheme is self-consistent. Power Divergence
Subtraction moves hence a somewhat arbitrary amount of the short distance
contributions from loops to counterterms and makes precise cancellations
manifest which arise from fine tuning. Notice that the re-summed deuteron
propagator has the same order $1/(MQ)$ as each diagram in
Fig.~\ref{fig:deuteronprop}.

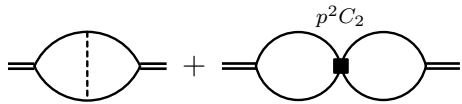
\begin{figure}[!htb]
  \begin{center}
    \feynbox{60\unitlength}{
            \begin{fmfgraph*}(60,30)
              \fmfleft{i} \fmfright{o} \fmf{double,width=thin,tension=8}{i,v1}
              \fmf{double,width=thin,tension=8}{v2,o}
              \fmf{vanilla,width=thin,left=0.65}{v1,v2}
              \fmf{vanilla,width=thin,left=0.65}{v2,v1} \fmffreeze
              \fmffreeze \fmfipath{pa} \fmfiset{pa}{vpath(__v1,__v2)}
              \fmfipath{pb} \fmfiset{pb}{vpath(__v2,__v1)} \fmfi{dashes}{point
                1/2 length(pa) of pa -- point 1/2 length(pb) of pb}
              \end{fmfgraph*}}\hq+\hq
            \feynbox{90\unitlength}{
            \begin{fmfgraph*}(90,40)
              \fmfleft{i} \fmfright{o} \fmf{double,width=thin,tension=5}{i,v1}
              \fmf{double,width=thin,tension=5}{v2,o}
              \fmf{vanilla,width=thin,left=0.8}{v1,v3}
              \fmf{vanilla,width=thin,left=0.8}{v3,v1}
              \fmf{vanilla,width=thin,left=0.8}{v2,v3}
              \fmf{vanilla,width=thin,left=0.8}{v3,v2}
              \fmfv{decor.shape=square,decor.size=5,label=$\fs p^2C_2$,
                label.angle=90,label.dist=0.18w}{v3}
            \end{fmfgraph*}}
          \vspace{10pt}
    \caption{The NLO corrections to the deuteron.}
    \label{fig:NLOcorrections}
  \end{center}
\end{figure}

One surprising result arises from this analysis because chiral symmetry
implies a derivative coupling of the pion to the nucleon at leading order.
The contribution from one pion exchange includes a factor of $Q^{-2}$ from the
pion propagator and a factor of $Q^2$ coming from the pion-nucleon vertices,
so that for momenta of the order of the pion mass, the instantaneous one pion
exchange scales as $Q^0$ and is {\it smaller} than the contact piece $C_0$
which according to (\ref{scalingksw}) scales as $Q^{-1}$.  Iterated and
radiative pion exchanges are suppressed even further. Pion exchange and higher
derivative contact terms appear hence only as perturbations at higher orders.
In contradistinction to iterative potential model approaches, each higher
order contribution is inserted only once. In this scheme, the only
non-perturbative physics responsible for nuclear binding is extremely simple,
and the more complicated pion contributions are at each order given by a
finite number of diagrams. For example, the NLO contributions to the deuteron
are the one instantaneous pion exchange and the four nucleon interaction with
two derivatives, Fig.~\ref{fig:NLOcorrections}. The constants are
determined e.g.~by demanding the correct deuteron pole position and
residue~\cite{PRS}.

\noindent
In the two body sector, the theory thus emerging has been put to extensive
tests at NLO and NNLO, giving for the first time analytic answers to many
deuteron properties, see e.g.~\cite{INTWorkshopSummary}.  Although in general
process dependent, the expansion parameter is found to be of the order of
$\frac{1}{3}$ in most applications, so that NLO calculations can be expected
to be accurate to about $10\%$, and NNLO calculations to about $4\%$.  In all
cases, experimental agreement is within the estimated theoretical
uncertainties, and in some cases, previously unknown counterterms could be
determined.

\begin{figure}[!htb] 
  \centerline{\epsfig{file=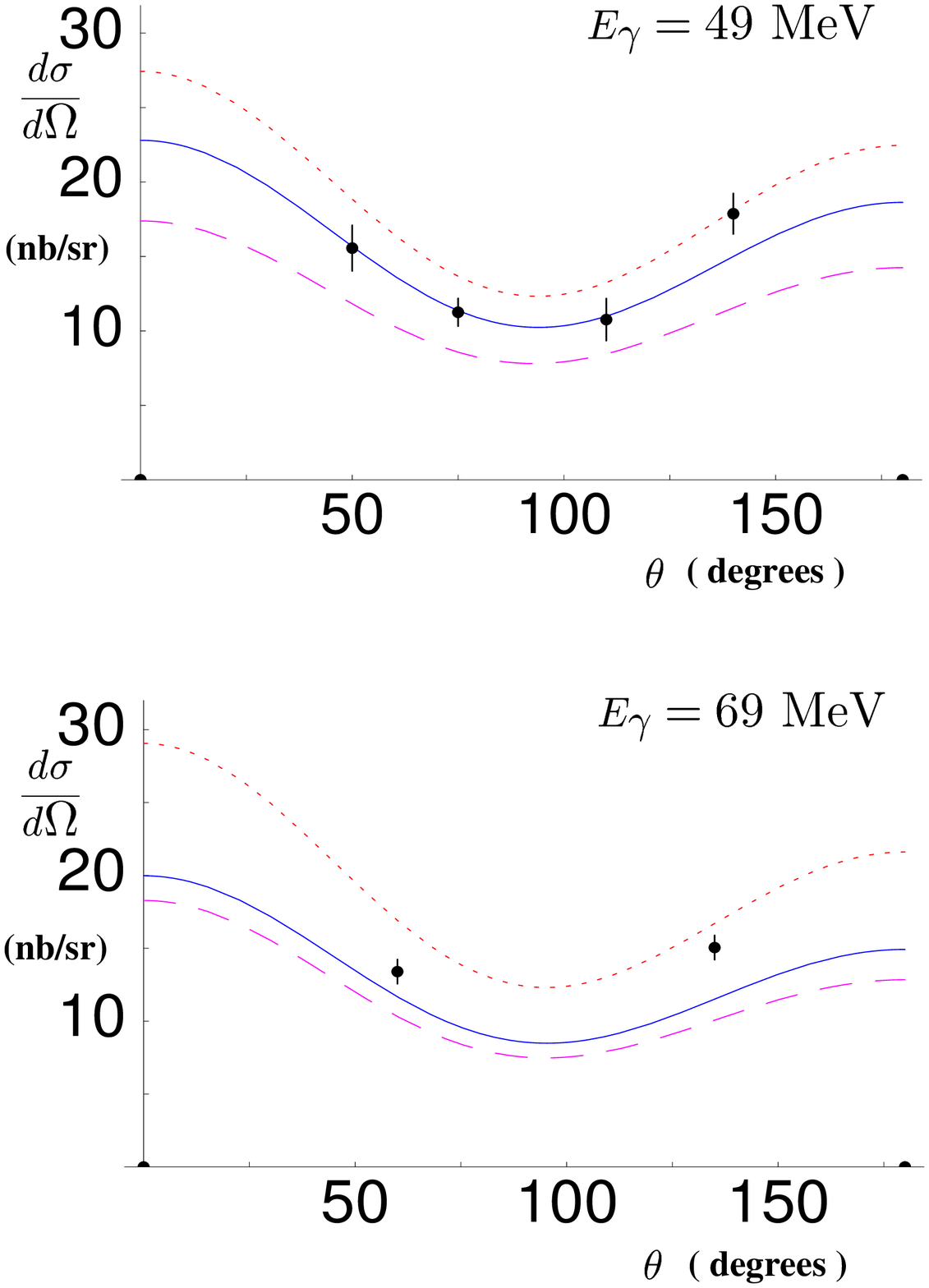,width=0.46\textwidth,
      bburx=484,bbury=704,bbllx=0,bblly=409,clip=} \hspace{0.05\textwidth}
    \epsfig{file=angdist4969_b.eps,width=0.46\textwidth,
      bburx=484,bbury=358,bbllx=0,bblly=62,clip=}} \vspace{10pt}
\caption{The differential cross section for elastic 
  \protect$\gamma$-deuteron Compton scattering at incident photon energies of
  \protect$E_{\gamma}=49\ {\rm MeV}$ and \protect$69\ {\rm MeV}$ in an EFT
  with explicit pions~\protect\cite{Compton}, no free parameters. Dashed: LO;
  dotted: NLO without the graphs that contribute to the nucleon
  polarisability; solid curve: complete NLO result.}
\label{fig:compton}
\end{figure}

The elastic deuteron Compton scattering diagrams to NLO are partially obtained
by gauging the Lagrangean (\ref{ksw}), i.e.~by replacing ordinary derivatives
by covariant ones: At LO, a seagull-graph and one graph in which the incident
and outgoing photon couple to the same nucleon are found. At NLO, the photons
are attached in all possible ways to the corrections in
Fig.~\ref{fig:NLOcorrections}, including to the pion, the $NN\pi$ vertex and
the $C_2$ vertex. The Fermi interaction $\vec{\sigma}\cdot\vec{B}$ probes the
${}^1{\rm S}_0$ intermediate $NN$ state and enters at NLO, too.  Finally, the
iso-scalar electric nuclear polarisability was shown to come from relativistic
(``radiative'') pions in Chiral Perturbation Theory~\cite{BKMa},
$\alpha_{E,N}=\frac{5g_A^2\alpha}{48\pi{\fpi^2}\mpi}$, and is NLO. The cross
section fits finally on less than one page with functions not more complicated
than Logarithms and Arcustangentes~\cite{Compton} and contains no free
parameters. Comparison with the Urbana experiment~\cite{Lucas} in
Fig.~\ref{fig:compton} shows good agreement, with the pion graphs that
dominate the electric polarisability of the nucleon necessary to improve it.
The deuteron scalar and tensor electric and magnetic polarisabilities are also
easily extracted~\cite{pola}.

\noindent In the three body sector, even the leading order calculation is too
complex for a fully analytical solution. Still, the equations that need to be
solved are computationally trivial and can furthermore be improved
systematically by higher order corrections that involve only (partly
analytical, partly numerical) integrations, as opposed to many-dimensional
integral equations arising in other approaches. The $nd$ system provides a
laboratory in which many complications of the other channels are not
encountered: The absence of Coulomb interactions ensures that only properties
of the strong interactions are probed. In the quartet channel, the Pauli
principle forbids three body forces~\cite{Stooges} in the first few orders.
Because the calculation is parameter-free, it allows one to determine the
range of validity of the KSW scheme without a detailed analysis of the fitting
procedure. Although e.g.~the quartet scattering length is large, no extra fine
tuning except the one for the deuteron is required. In the ${\rm S}$ wave,
spin-doublet (triton) channel, the situation is more complicated. An unusual
renormalisation of the three-body force makes it large and as important as the
leading two-body forces~\cite{Stooges2}. More work is needed there.

\begin{figure}[!htb]
  \begin{center}
    \setlength{\unitlength}{0.7pt}
    \feynbox{50\unitlength}{
            \begin{fmfgraph*}(50,50)
              \fmfleft{i2,i1} \fmfright{o2,o1}
              \fmf{double,tension=6}{i1,v1,v2} \fmf{vanilla,width=thin}{v2,o1}
              \fmf{double,tension=6}{v3,v4,o2} \fmf{vanilla,width=thin}{i2,v3}
              \fmffreeze \fmf{vanilla,width=thin}{v2,v3}
            \end{fmfgraph*}}
          \hqm$+$\hqm \feynbox{80\unitlength}{
            \begin{fmfgraph*}(80,50)
              \fmfleft{i2,i1} \fmfright{o2,o1} \fmf{double,tension=4}{i1,v1}
              \fmf{vanilla,width=thin}{v1,v2} \fmf{double,tension=4}{o1,v2}
              \fmf{vanilla,width=thin}{i2,v3} \fmf{vanilla,width=thin}{v3,o2}
              \fmffreeze \fmf{vanilla,width=thin}{v2,v3}
              \fmf{vanilla,width=thin}{v3,v1}
            \end{fmfgraph*}}
          \hqm$+$\hqm \feynbox{110\unitlength}{
            \begin{fmfgraph*}(110,50)
              \fmfleft{i2,i1} \fmfright{o2,o1} \fmf{double,tension=8}{i1,v1}
              \fmf{vanilla,width=thin}{v1,v2} \fmf{double,tension=8}{o1,v2}
              \fmf{double,tension=100}{v3,v3a}
              \fmf{double,tension=100}{v4,v4a} \fmf{phantom}{v3a,v4a}
              \fmf{vanilla,width=thin}{i2,v3} \fmf{vanilla,width=thin}{v4,o2}
              \fmffreeze \fmf{vanilla,width=thin}{v2,v4}
              \fmf{vanilla,width=thin}{v3,v1} \fmffreeze
              \fmf{vanilla,width=thin,left=0.65}{v3a,v4a}
              \fmf{vanilla,width=thin,left=0.65}{v4a,v3a}
            \end{fmfgraph*}}
          \hqm$+$\hqm \feynbox{110\unitlength}{
            \begin{fmfgraph*}(110,50)
              \fmfleft{i2,i1} \fmfright{o2,o1} \fmf{double,tension=8}{i1,v1}
              \fmf{vanilla,width=thin}{v1,v4} \fmf{double,tension=100}{v4,v5}
              \fmf{vanilla,width=thin,tension=1.666}{v5,o1}
              \fmf{double,tension=8}{o2,v6} \fmf{vanilla,width=thin}{v6,v3}
              \fmf{double,tension=100}{v3,v2}
              \fmf{vanilla,width=thin,tension=1.666}{v2,i2} \fmffreeze
              \fmf{vanilla,width=thin}{v1,v2} \fmf{vanilla,width=thin}{v3,v4}
              \fmf{vanilla,width=thin}{v5,v6}
            \end{fmfgraph*}}
          \hqm$+\;\dots$
          \\[5ex]
          \feynbox{90\unitlength}{
            \begin{fmfgraph*}(90,50)
              \fmfleft{i2,i1} \fmfright{o2,o1}
              \fmf{double,width=thin,tension=3}{i1,v1}
              \fmf{double,width=thin,tension=1.5}{v1,v3,v2}
              \fmf{double,width=thin,tension=3}{v2,o1}
              \fmf{vanilla,width=thin}{i2,v4,o2} \fmffreeze \fmffreeze
              \fmf{ellipse,rubout=1}{v3,v4}
              \end{fmfgraph*}}
            \hq$=$\hq \feynbox{90\unitlength}{
            \begin{fmfgraph*}(90,50)
              \fmfleft{i2,i1} \fmfright{o2,o1}
              \fmf{double,width=thin,tension=4}{i1,v1,v2}
              \fmf{vanilla,width=thin}{v2,o1}
              \fmf{double,width=thin,tension=4}{v3,v4,o2}
              \fmf{vanilla,width=thin}{i2,v3} \fmffreeze
              \fmf{vanilla,width=thin}{v2,v3}
              \end{fmfgraph*}}
            \hq$+$\hq \feynbox{170\unitlength}{
            \begin{fmfgraph*}(170,50)
              \fmfleft{i2,i1} \fmfright{o2,o1}
              \fmf{double,width=thin,tension=3}{i1,v1}
              \fmf{double,width=thin,tension=1.5}{v1,v6,v5}
              \fmf{double,width=thin,tension=3}{v5,v2}
              \fmf{vanilla,width=thin}{v2,o1} \fmf{vanilla,width=thin}{i2,v7}
              \fmf{vanilla,width=thin,tension=0.666}{v7,v4}
              \fmf{double,width=thin,tension=4}{v4,v3,o2} \fmffreeze
              \fmf{vanilla,width=thin}{v4,v2} \fmf{ellipse,rubout=1}{v6,v7}
              \end{fmfgraph*}}
            \vspace{10pt}
    \caption{The double infinite series of LO
      ``pinball'' diagrams, some of which are shown in the first line, is
      equivalent to the solution of the Faddeev equation shown in the second
      line.}
    \label{fig:LOfaddeev}
  \end{center}
\end{figure}
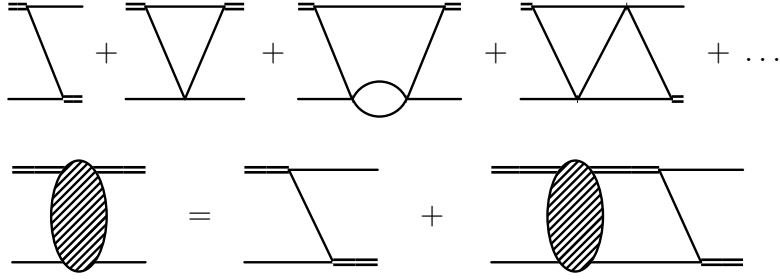

A comparative study between the theory with explicit pions and the one with
pions integrated out was performed in~\cite{pbhg} for the spin quartet ${\rm
  S}$ wave. As seen above, the two theories are identical at LO: All graphs
involving only $C_0$ interactions are of the same order and form a double
series which is not geometrical and cannot be summed analytically. One is
hence left with the task of summing all ``pinball'' diagrams (first line of
Fig.~\ref{fig:LOfaddeev}).  Summing all ``bubble-chain'' sub-graphs into the
deuteron propagator, one can however obtain the solution numerically from the
integral equation pictorially shown in the lower line of
Fig~\ref{fig:LOfaddeev}. A code runs within seconds on a personal computer.

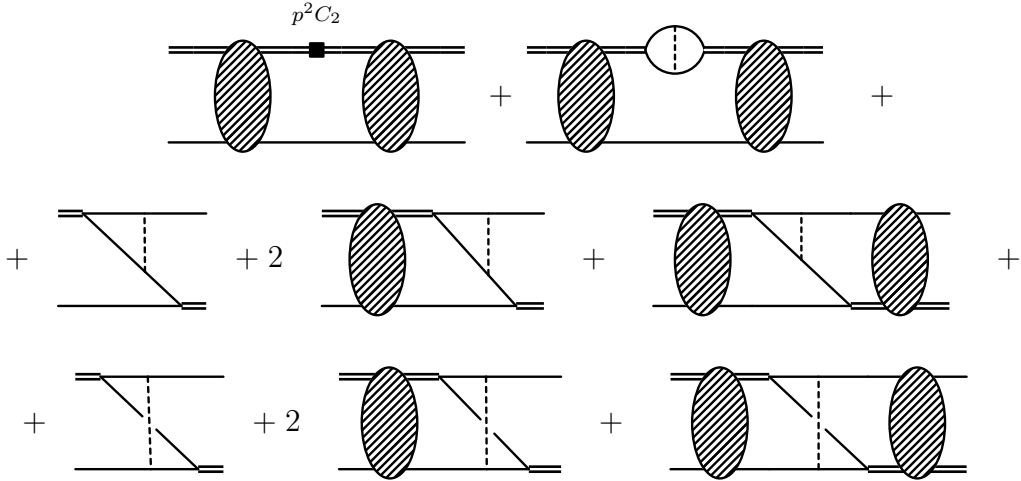
\begin{figure}[!htb]
  \begin{center}
    \setlength{\unitlength}{0.7pt}
%
%
    \feynbox{200\unitlength}{
            \begin{fmfgraph*}(200,50)
              \fmfleft{i2,i1} \fmfright{o2,o1}
              \fmf{double,width=thin,tension=3}{i1,v1}
              \fmf{double,width=thin,tension=1.5}{v1,v3,v2}
              \fmf{double,width=thin,tension=3}{v2,v5}
              \fmfv{decor.shape=square,decor.size=5,label=$\fs p^2
                C_2$,label.angle=90,label.dist=0.08w}{v5}
              \fmf{double,width=thin,tension=3}{v5,v7}
              \fmf{double,width=thin,tension=1.5}{v7,v8,v9}
              \fmf{double,width=thin,tension=3}{v9,o1}
              \fmf{vanilla,width=thin}{i2,v4}
              \fmf{vanilla,width=thin,tension=0.5}{v4,v10}
              \fmf{vanilla,width=thin}{v10,o2} \fmffreeze \fmffreeze
              \fmf{ellipse,rubout=1}{v3,v4} \fmf{ellipse,rubout=1}{v8,v10}
              \fmf{vanilla,width=thin}{v3,v4} \fmf{vanilla,width=thin}{v8,v10}
              \end{fmfgraph*}}
            \hqm\hqm\hqm$+$\hqm\hqm\hqm \feynbox{200\unitlength}{
            \begin{fmfgraph*}(200,50)
              \fmfleft{i2,i1} \fmfright{o2,o1}
              \fmf{double,width=thin,tension=3}{i1,v1}
              \fmf{double,width=thin,tension=1.5}{v1,v3,v2}
              \fmf{double,width=thin,tension=3}{v2,v5}
              \fmf{vanilla,width=thin,left=0.8,tension=0.5}{v5,v6}
              \fmf{vanilla,width=thin,left=0.8,tension=0.5}{v6,v5}
              \fmf{double,width=thin,tension=3}{v6,v7}
              \fmf{double,width=thin,tension=1.5}{v7,v8,v9}
              \fmf{double,width=thin,tension=3}{v9,o1}
              \fmf{vanilla,width=thin}{i2,v4}
              \fmf{vanilla,width=thin,tension=0.333}{v4,v10}
              \fmf{vanilla,width=thin}{v10,o2} \fmffreeze \fmffreeze
              \fmf{ellipse,rubout=1}{v3,v4} \fmf{ellipse,rubout=1}{v8,v10}
              \fmf{vanilla,width=thin}{v3,v4} \fmf{vanilla,width=thin}{v8,v10}
              \fmffreeze \fmfipath{pa} \fmfiset{pa}{vpath(__v5,__v6)}
              \fmfipath{pb} \fmfiset{pb}{vpath(__v6,__v5)} \fmfi{dashes}{point
                1/2 length(pa) of pa -- point 1/2 length(pb) of pb}
              \end{fmfgraph*}}
            $+$
            
            \vspace*{5ex}

            
            $+$ \feynbox{100\unitlength}{
            \begin{fmfgraph*}(100,50)
              \fmfleft{i2,i1} \fmfright{o2,o1}
              \fmf{double,width=thin,tension=2.5}{i1,v1}
              \fmf{vanilla,width=thin}{v1,v3,o1}
              \fmf{double,width=thin,tension=5}{o2,v2}
              \fmf{vanilla,width=thin}{v2,i2} \fmffreeze
              \fmf{vanilla,width=thin,tension=1.7}{v2,v4}
              \fmf{vanilla,width=thin}{v4,v1} \fmffreeze
              \fmf{dashes,width=thin}{v4,v3}
              \end{fmfgraph*}}
            $+\;2$ \feynbox{150\unitlength}{
            \begin{fmfgraph*}(150,50)
              \fmfleft{i2,i1} \fmfright{o2,o1}
              \fmf{double,width=thin}{i1,i1a,v1}
              \fmf{vanilla,width=thin}{v1,v3,o1}
              \fmf{double,width=thin,tension=5}{o2,v2}
              \fmf{vanilla,width=thin}{v2,i2a}
              \fmf{vanilla,width=thin,tension=2.5}{i2a,i2} \fmffreeze
              \fmf{vanilla,width=thin,tension=2}{v2,v4}
              \fmf{vanilla,width=thin}{v4,v1} \fmffreeze
              \fmf{dashes,width=thin}{v4,v3} \fmffreeze
              \fmf{ellipse,rubout=1}{i1a,i2a} \fmf{vanilla}{i1a,i2a}
              \end{fmfgraph*}}
            $+$ \feynbox{200\unitlength}{
            \begin{fmfgraph*}(200,50)
              \fmfleft{i2,i1} \fmfright{o2,o1}
              \fmf{double,width=thin}{i1,i1a,v1}
              \fmf{vanilla,width=thin}{v1,v1a,v3}
              \fmf{vanilla,width=thin}{v3,o1a,o1}
              \fmf{double,width=thin}{v4,o2a,o2}
              \fmf{vanilla,width=thin,tension=0.5}{v4,v2}
              \fmf{vanilla,width=thin}{v2,i2a,i2} \fmffreeze
              \fmf{vanilla,width=thin}{v1,v5} \fmf{vanilla,width=thin}{v5,v4}
              \fmffreeze \fmf{dashes,width=thin}{v5,v1a} \fmffreeze
              \fmf{ellipse,rubout=1 }{i1a,i2a} \fmf{vanilla}{i1a,i2a}
              \fmf{ellipse,rubout=1 }{o1a,o2a} \fmf{vanilla}{o1a,o2a}
              \end{fmfgraph*}}
            $+$
            
            \vspace{5ex}

            
            $+$ \feynbox{100\unitlength}{
            \begin{fmfgraph*}(100,50)
              \fmfleft{i2,i1} \fmfright{o2,o1}
              \fmf{double,width=thin,tension=3}{i1,v1}
              \fmf{vanilla,width=thin,tension=1.6}{v1,v3}
              \fmf{vanilla,width=thin}{v3,o1}
              \fmf{double,width=thin,tension=3}{o2,v2}
              \fmf{vanilla,width=thin,tension=1.6}{v2,v5}
              \fmf{vanilla,width=thin}{v5,i2} \fmffreeze
              \fmf{vanilla,width=thin}{v2,v4} \fmf{phantom,tension=3}{v4,v4a}
              \fmf{vanilla,width=thin}{v4a,v1} \fmffreeze
              \fmf{dashes,width=thin}{v3,v5}
              \end{fmfgraph*}}         
            $+\;2$ \feynbox{150\unitlength}{
            \begin{fmfgraph*}(150,50)
              \fmfleft{i2,i1} \fmfright{o2,o1}
              \fmf{double,width=thin,tension=1.5}{i1,i1a,v1}
              \fmf{vanilla,width=thin,tension=1.6}{v1,v3}
              \fmf{vanilla,width=thin}{v3,o1}
              \fmf{double,width=thin,tension=3}{o2,v2}
              \fmf{vanilla,width=thin,tension=2.3}{v2,v5}
              \fmf{vanilla,width=thin}{v5,i2a}
              \fmf{vanilla,width=thin,tension=1.9}{i2a,i2} \fmffreeze
              \fmf{dashes,width=thin}{v3,v5} \fmf{ellipse,rubout=1 }{i1a,i2a}
              \fmf{vanilla}{i1a,i2a} \fmffreeze
              \fmf{vanilla,width=thin,tension=1.2}{v2,v4}
              \fmf{phantom,tension=3}{v4,v4a} \fmf{vanilla,width=thin}{v4a,v1}
              \end{fmfgraph*}}
            $+$ \feynbox{200\unitlength}{
            \begin{fmfgraph*}(200,50)
              \fmfleft{i2,i1} \fmfright{o2,o1}
              \fmf{double,width=thin}{i1,i1a,v1}
              \fmf{vanilla,width=thin}{v1,v1a,v3}
              \fmf{vanilla,width=thin}{v3,o1a,o1}
              \fmf{double,width=thin}{v4,o2a,o2}
              \fmf{vanilla,width=thin}{v4,v2a,v2}
              \fmf{vanilla,width=thin}{v2,i2a,i2} \fmffreeze
              \fmf{vanilla,width=thin}{v1,v5a}
              \fmf{phantom,width=thin,tension=3}{v5b,v5a}
              \fmf{vanilla,width=thin}{v5b,v4} \fmffreeze
              \fmf{dashes,width=thin}{v2a,v1a} \fmffreeze
              \fmf{ellipse,rubout=1 }{i1a,i2a} \fmf{vanilla}{i1a,i2a}
              \fmf{ellipse,rubout=1 }{o1a,o2a} \fmf{vanilla}{o1a,o2a}
              \end{fmfgraph*}}
  \end{center}
            \vspace{10pt}
    \caption{The NLO contributions to $nd$ scattering in the quartet channel.
      First line: Corrections to the deuteron propagator; second line: pionic
      corrections to the $C_0$ vertex ; third line: pionic corrections to
      three particle breakup in the intermediate state. Permuted graphs left
      out.}
    \label{fig:NLOthreebody}  
\end{figure}

The power counting shows that at NLO, we have additional contributions from:
$p^2C_2$ insertions and pion exchange corrections to the deuteron propagator
depicted in the first line of Fig.~\ref{fig:NLOthreebody}; pionic vertex
corrections to $C_0$ (second line); and the pion diagram of the last line
which corrects the three particle intermediate state. Here, we used a
re-formulation of the Lagrangean (\ref{ksw}) in order not to have poorly
convergent diagrams containing $C_2$ like the second one in the first line of
Fig.~\ref{fig:LOfaddeev}, see~\cite{pbhg} for details. The calculation without
explicit pions was carried out to NNLO.

\begin{figure}[!htb] 
  \centerline{\epsfig{file=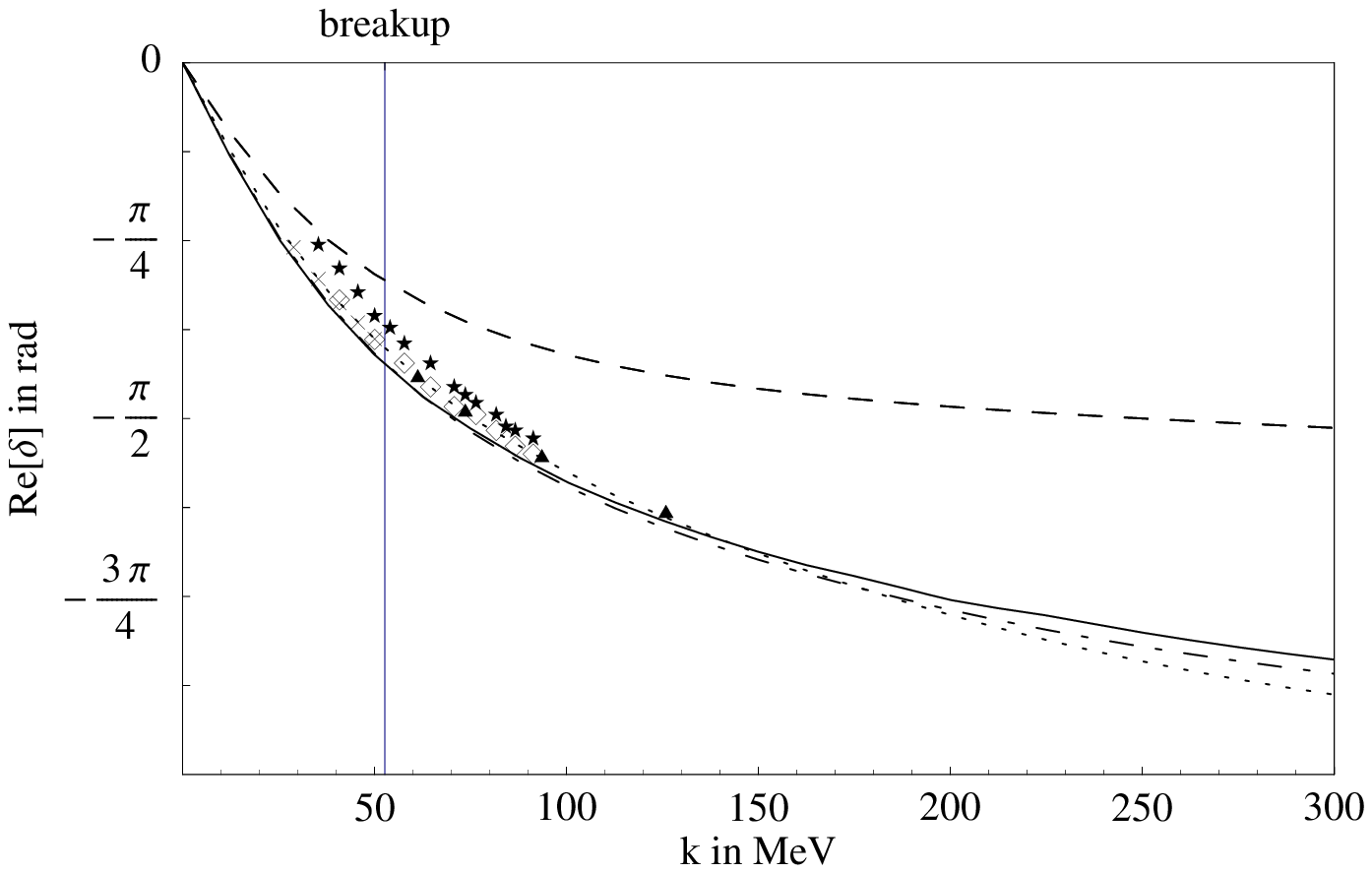,width=0.5\textwidth,clip=}
    \hspace{-1.5ex}
              \epsfig{file=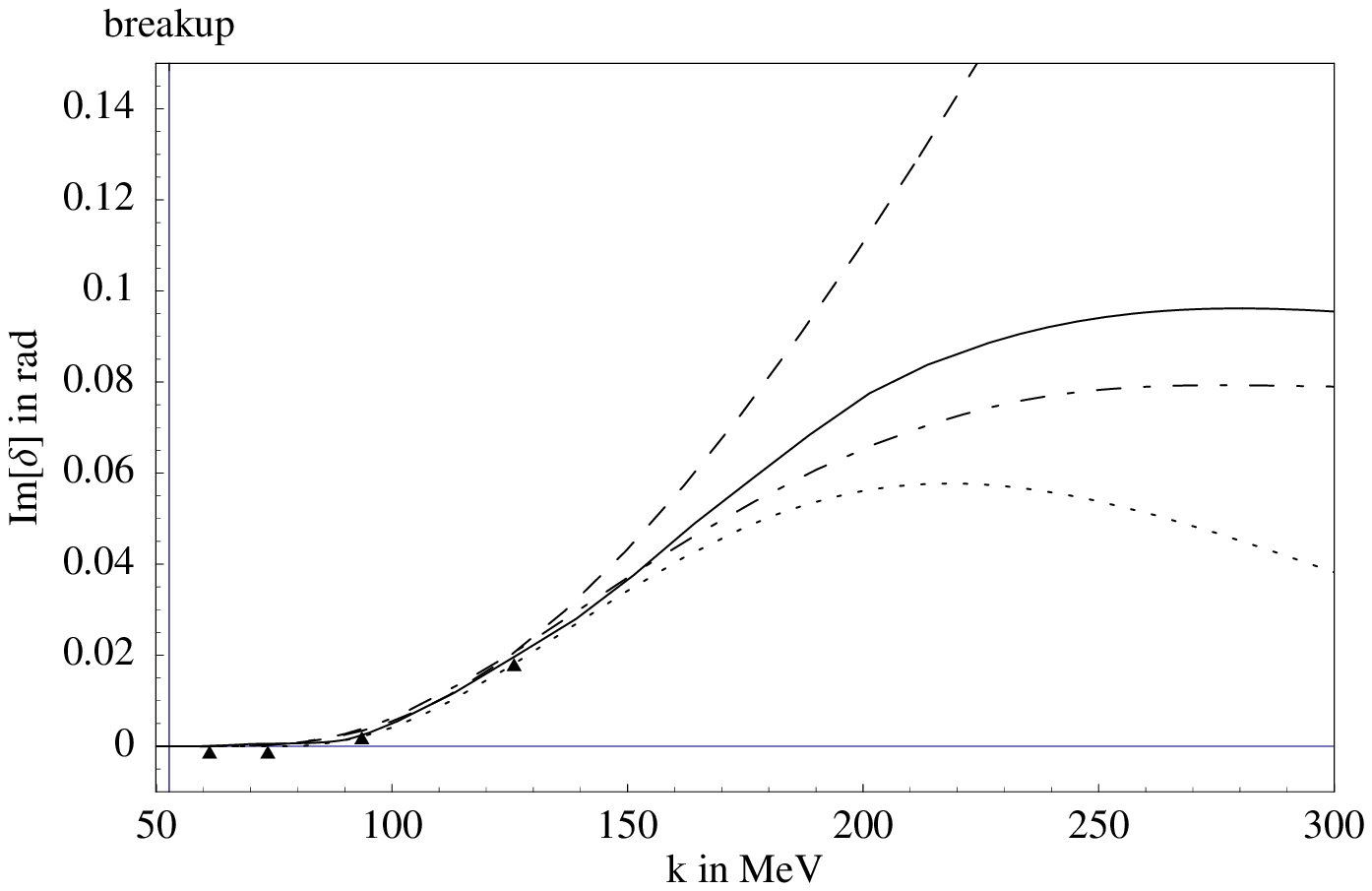,width=0.5\textwidth,clip=}}
  \vspace{10pt}
\caption{Real and imaginary parts in the quartet \protect${\rm S}$ wave
  phase shift of \protect$nd$ scattering versus the centre-of-mass
  momentum~\protect\cite{pbhg}. Dashed: LO; solid (dot-dashed) line: NLO with
  perturbative pions (pions integrated out); dotted: NNLO without
  pions~\protect\cite{Stooges,pbfghg}. Realistic potential
  models: squares from~\protect\cite{WitalaTUNL}, crosses
  from~\protect\cite{Kievsky}, triangles from~\protect\cite{Hueberetal}.
  Stars: TUNL \protect$pd$ phase shift analysis~\protect\cite{WitalaTUNL}.}
\label{fig:delta}
\end{figure}

All calculations demonstrate convergence. The scattering length is
$a({}^4S_\frac{3}{2},{\rm LO})=(5.1\pm 1.5)\;\fm$, and at NLO with (without)
perturbative pions $a({}^4S_\frac{3}{2},{\rm NLO,\pi})=(6.8\pm 0.7)\;\fm$
($a({}^4S_\frac{3}{2},{\rm NLO,{\rm no}\pi})=(6.7\pm 0.7)\;\fm$). At
NNLO,~\cite{Stooges} report $a({}^4S_\frac{3}{2},{\rm NNLO,{\rm
    no}\pi})=(6.33\pm 0.1)\;\fm$, and the experimental value is
$a({}^4S_\frac{3}{2},{\rm exp})=(6.35\pm 0.02)\;\fm$~\cite{Dilgetal}.
Comparing the NLO correction to the LO scattering length provides one with the
familiar error estimate at NLO: $(\frac{1}{3})^2\approx 10\%$. The NLO
calculations with and without pions lie within each other's error bar. The
NNLO calculation is inside the error ascertained to the NLO calculation and
carries itself an error of about $(\frac{1}{3})^3\approx 4\%$.  NLO and LO
contributions become comparable for momenta of more than $200\;\MeV$. In the
imaginary part shown in Fig.~\ref{fig:delta}, the same pattern emerges with a
slightly more pronounced difference between the pion-less and pions-full
theory.  Because results obtained with EFT are easily dissected for the
relative importance of the various terms, one concludes that pionic
corrections to $nd$ scattering in the quartet ${\rm S}$ wave channel --
although formally NLO -- are indeed much weaker: The calculation with
perturbative pions and with pions integrated out do not differ significantly
over a wide range of momenta. The difference should appear for momenta of the
order of $\mpi$ and higher because of non-analytical contributions of the pion
cut.  However, it is very moderate for momenta of up to $300\;\MeV$ in the
centre-of-mass frame ($E_{{\rm cm}}\approx70\;\MeV$), see
Fig.~\ref{fig:delta}.  This and the lack of data makes it
difficult to assess whether the KSW power counting scheme to include pions as
perturbative increases the range of validity over the pion-less theory, but
effects from the pion cut are seemingly weak.

\begin{figure}[!htb] 
  \centerline{\epsfig{file=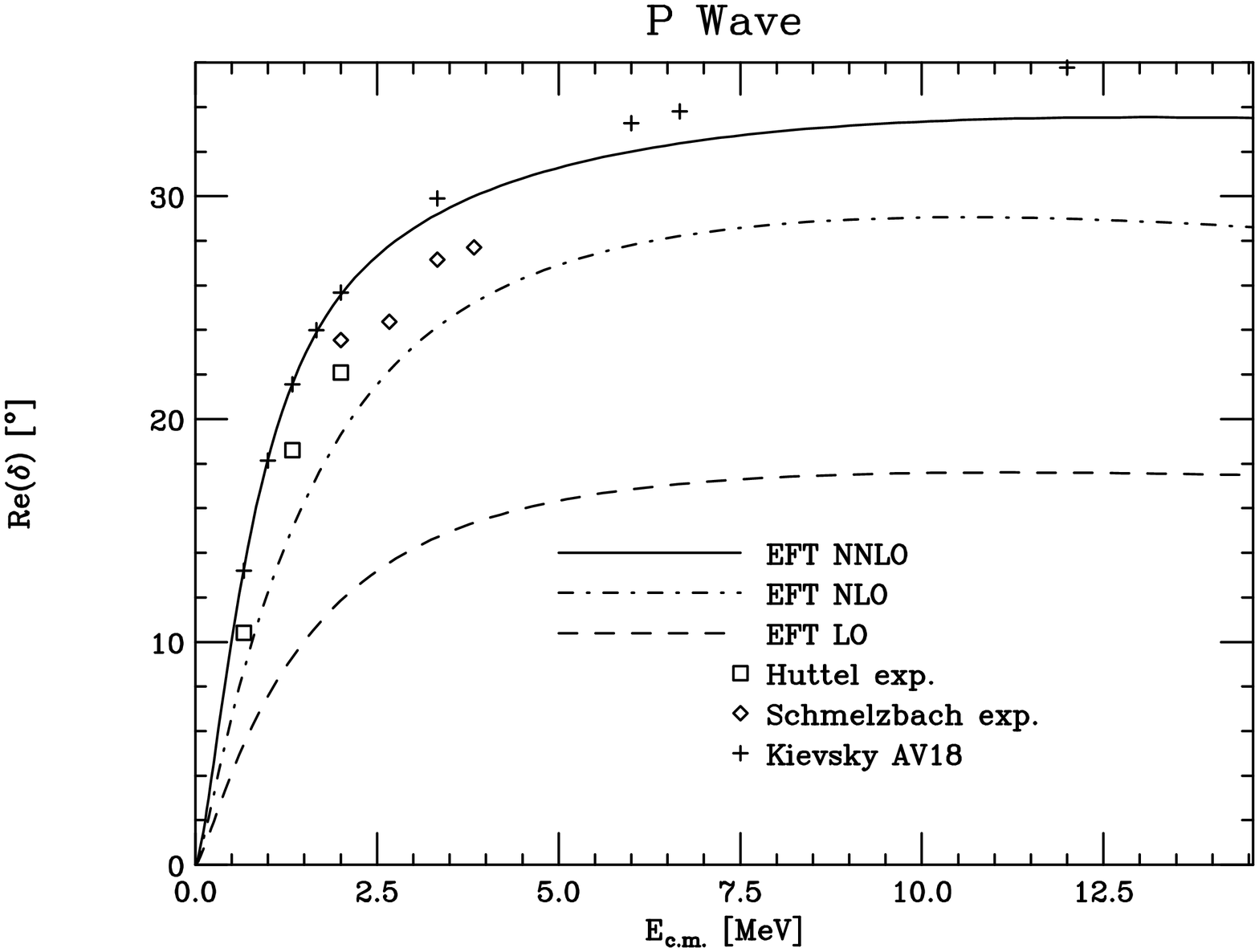,height=0.45\textwidth,angle=90}
    \hspace{0.05\textwidth}
    \epsfig{file=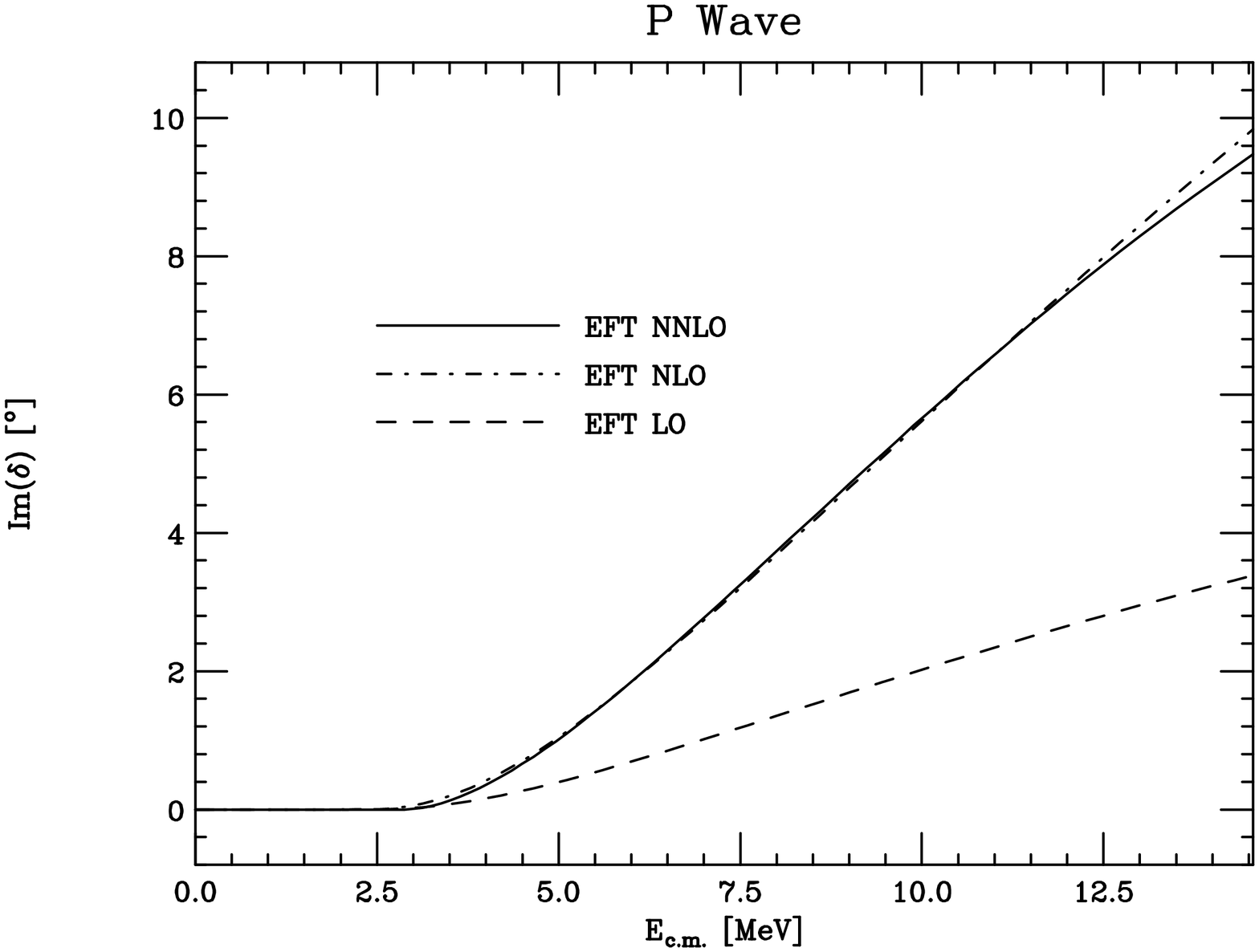,height=0.45\textwidth,angle=90}} \vspace{20pt}
  \centerline{\epsfig{file=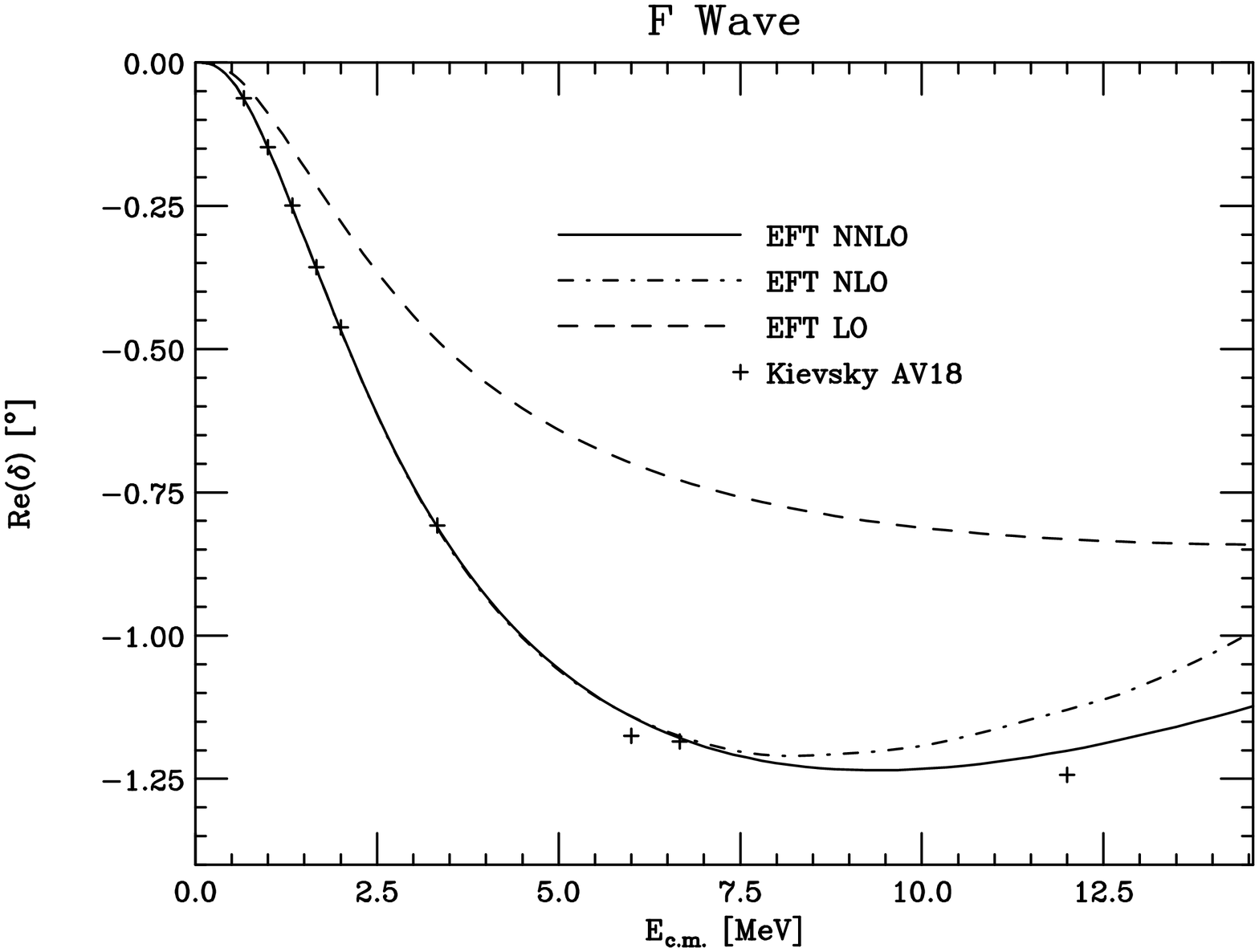,height=0.45\textwidth,angle=90}
    \hspace{0.05\textwidth}
    \epsfig{file=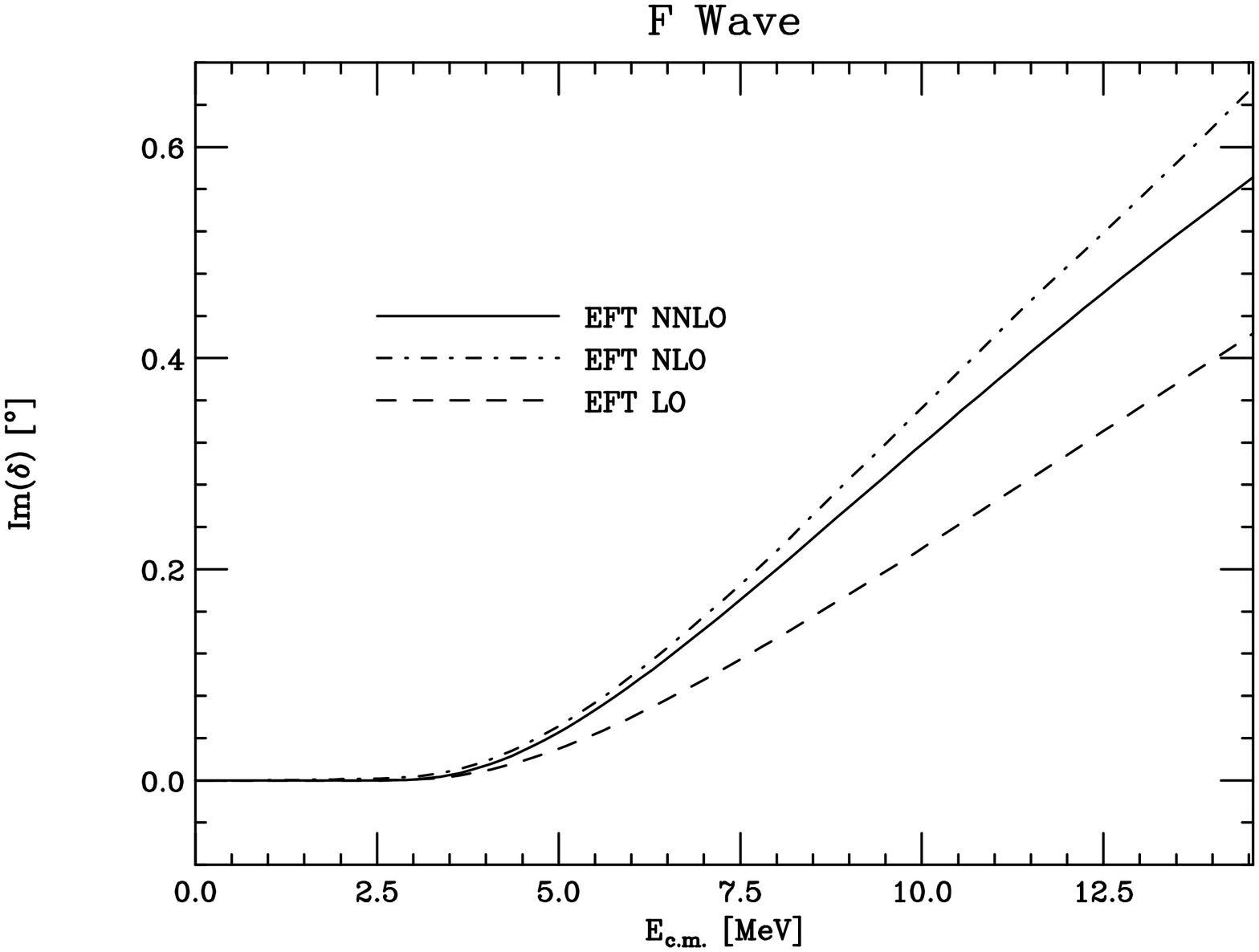,height=0.45\textwidth,angle=90}} \vspace{10pt}
\caption{Real and imaginary parts of the quartet \protect${\rm P}$ (top) and
  doublet \protect${\rm F}$ (bottom) wave phase shift of \protect$nd$
  scattering versus the centre-of-mass energy in the EFT without
  pions~\protect\cite{pbfghg}. Dashed: LO; dot-dashed: NLO; solid line: NNLO.
  The experiments by Huttel et al.~\protect\cite{Hutteletal} and Schmelzbach
  et al.~\protect\cite{Schmelzbachetal} are denoted by open squares and
  diamonds, respectively. The calculations of Kievsky et al.~(crosses) are
  from Refs.~\protect\cite{Kievsky} below breakup
  (\protect$E_{{\rm cm}}=B$) and \protect\cite{Kievskyprivcomm} above
  breakup.}
\label{fig:quartetpdoubletf}
\end{figure}

\noindent
Finally, the real and imaginary parts of the higher partial waves
$l=1,\dots,4$ in the spin quartet and doublet channel were presented
in~\cite{pbfghg} in a papameter-free calculation.
Figure~\ref{fig:quartetpdoubletf} shows two examples. Comparison of the LO
with the NLO and NNLO result demonstrates convergence of the EFT, with the
expansion parameter again about $\frac{1}{3}$. It is interesting that the NLO
correction is for high enough energies sometimes sizeable, while the NNLO
correction is in general very small.
Within the range of validity of this pion-less theory, convergence is good,
and the results agree with potential model calculations (as available) within
the theoretical uncertainty. That makes one optimistic about carrying out
higher order calculations of problematic spin observables like the $A_y$
problem where the EFT approach will differ from potential model calculations
due to the inclusion of three-body forces.


\absatz {\textbf{Acknowledgements}}

\vspace{1ex}
\noindent It is a great pleasure to thank my collaborators -- J.-W.~Chen,
R.P.~Springer and M.J.~Savage in \cite{pola,Compton}, P.F.~Bedaque in
\cite{pbhg,pbfghg}, and F.~Gabbiani in \cite{pbfghg} -- for a lot of fun, and
the EFT group at the INT and the University of Washington
in Seattle for a number of valuable discussions. The work was supported in
part by the Department of Energy grant DE-FG03-97ER41014 and the
Bundesministerium f{\"u}r Bildung und Forschung.


\end{fmffile}
\end{document}